\documentclass[12pt,a4paper]{article}
\usepackage[T2A]{fontenc}
\usepackage[utf8]{inputenc}
\usepackage[english]{babel}
\usepackage{amssymb}
\usepackage{amsmath}
\usepackage{graphicx}
\usepackage[small,sc,center]{titlesec}
\usepackage{indentfirst}
\usepackage[labelsep=period,labelfont=bf,small]{caption}    %
\usepackage{array, longtable, afterpage}
\usepackage{changes}

\usepackage{floatrow} 
\usepackage[small,sc,center]{titlesec}
\usepackage{xcolor}

\usepackage{authblk}
\usepackage{enumitem} 
\usepackage{hyperref}
\usepackage{upgreek}
\usepackage{changepage}
\usepackage{microtype}
\usepackage{natbib}

\def\gtrsim{\ {\raise-.5ex\hbox{$\buildrel>\over\sim$}}\ }

\def\apgt{\ {\raise-.5ex\hbox{$\buildrel>\over\sim$}}\ }
\def\aplt{\ {\raise-.5ex\hbox{$\buildrel<\over\sim$}}\ }

\newlength{\bibitemsep}
\newlength{\bibparskip}
\let\oldthebibliography\thebibliography
\renewcommand\thebibliography[1]{%
  \oldthebibliography{#1}%
  \setlength{\parskip}{\bibitemsep}%
  \setlength{\itemsep}{\bibparskip}%
}

\newcommand\aj{Astron. J.}%
\newcommand\araa{Annu. Rev. Astron. Astrophys.}%
\newcommand\apj{Astrophys. J.}%
\newcommand\apjs{Astrophys. J. Suppl. Ser.}%
\newcommand\aap{Astron. Astrophys.}%
\newcommand\mnras{Mon. Not.  RAS}%
\sloppypar
\begin{document}
\emergencystretch 2em

\title{\bf  Looking for Signs of Unresolved Binarity in the Continuum of LAMOST Stellar Spectra}

\author{
Mikhail Prokhorov$^{1,*}$,
Kefeng Tan$^2$,
Nikolay Samus$^{1,3}$,
Ali Luo$^2$,
Dana Kovaleva$^3$
Jingkun Zhao$^2$,
Yujuan Liu $^2$,
Pavel Kaygorodov$^3$,
Oleg Malkov$^3$,
Sergey Sichevskij$^3$,
Lev Yungelson$^3$
Gang Zhao$^2$\\
\small{\begin{flushleft}
{\it $^1$Sternberg Astronomical Institute, M.V.~Lomonosov Moscow State University}\\
{\it $^2$National Astronomical Observatories, Chinese Academy of Sciences}\\
{\it $^3$Institute of Astronomy, Russian Academy of Sciences}
\end{flushleft}
}
}

\date{\small \today}

\maketitle
\textbf{Abstract} ---    
    We describe an attempt to derive the binarity rate of samples of  166
A-, F-, G-, and K-type stars from LAMOST DR5 and 1000 randomly selected 
presumably single stars from Gaia DR3 catalogs.  
To this end, we compared continua of the observed spectra with the continua of 
synthetic spectra from 
within 3700 $< \lambda < 9097$\,\AA.   
The latter spectra were reduced to the LAMOST set of wavelengths, while the 
former ones were smoothed. 
Next, we searched for every observed star of the nearest synthetic spectrum using a 
four-parameter representation---$T_{\rm eff}$, $\log g$, $[\textrm{Fe/H}]$, and a range 
of interstellar absorption values. 
However, rms deviations of observed spectra from
synthetic ones appeared to be not sufficient to claim that any of the stars is
a binary. 
We conclude that comparison of the intensity of pairs of spectral lines
remains the best way to detect binarity. 

\vfill
\noindent\rule{8cm}{1pt}\\
$^*$mike@sao.msu.ru

\section{Introduction}
     \label{sec:intro}

Binary stars are very numerous, and~binarity rate estimates increase from
\mbox{$\simeq$(20--30)\%} for M-dwarfs to $\simeq$(60--70)\% for A-type ones, while 
extreme ones for O stars reach  \mbox{(94 $\pm$ 14)\%}
\footnote{An estimate exceeding 100\% means that binaries are components of triples.} \citep{2013ARA&A..51..269D, 2017ApJS..230...15M, 2024BSRSL..93..170M, 2025A&A...693A.228C}. 
The importance of correct estimation of binarity rate is due to the role of binaries in formation of many extreme astrophysical objects, e.g.,~Supernovae, X-ray sources, gravitational wave sources, as~well as formation of populations of stars of certain types, for~instance, cataclysmic variables. No less important is the significance of the estimates of binarity rate for understanding star formation, dynamics of stellar aggregates, etc.  
(see, e.g., \citet{1981NInfo..49....3T, 1991MNRAS.251..293K, 2001MNRAS.322..231K, 2018MNRAS.473.5043E, 2019ApJ...874..127B, 2021ApJ...912..137S}).

Extensive sets of observational data for the stellar population of the Galaxy became available 
recently, as~a result of several surveys. Among~the most notable ones are the astrometric survey Gaia \citep{2016A&A...595A...1G, 2023A&A...674A...1G}, multi-band survey SDSS \citep{2023ApJS..267...44A},
and spectroscopic surveys (see below).
One may expect that a large proportion of the objects in these surveys are unresolved binary~stars. 

There are a number of approaches for resolving hidden binaries and  assessing their characteristics, which employ different samples of observable characteristics and differ in the scope of physical and observable parameters of unresolved binaries that they may~recover.

For the Gaia sources, the~methods of detecting probable unresolved 
binaries based on Gaia supplementary 
information about astrometric solutions 
are proposed \citep{2024arXiv240414127C, 2022MNRAS.513.5270P}. 
The methods based solely on photometric information \citep{2023AJ....165...45M, 2015BaltA..24..137C} require a combined multicolor photometry. 
A variety of methods use a combination of photometric information and trigonometric parallaxes \citep{2024MNRAS.527.8718W, 2021ApJS..253...22X, 1999ARep...43...80K}.
A combination of the data from several surveys may also be
helpful for finding unresolved binaries. 
Each of these methods is able to reveal a certain fraction of the population of unresolved binary stars using the multidimensional space of observational parameters.
Unfortunately, no one of them is universal, so they may be considered as~intercomplementary. 

Along with astrometric and photometric surveys, spectroscopic surveys,
such as LAMOST \citep{2015RAA....15.1095L}, RAVE \citep{McMillan2020}, GALAH \citep{2021MNRAS.506..150B},
APOGEE \citep{Joensson2020, 2022ApJS..259...35A}, and GES \citep{2012Msngr.147...25G, 2013Msngr.154...47R},
are an invaluable provider of information about unresolved binary systems.
Spectroscopic binaries are, on~the one hand, a~fairly representative observational class of objects,
and on the other hand, the~parameters obtained from solving the spectroscopic orbit
include almost all the orbital elements and the astrophysical parameters of the components.
Compared with visual binaries, spectroscopic binaries
have relatively short orbital periods. With~this property, it is
easier to construct their full orbital solutions by observing 
two stars orbiting each other multiple times.
Spectroscopic searches make it possible to 
apprehend how unresolved binarity affects spectra of the sources 
\citep{2025arXiv250114494L, 2024MNRAS.530.1935S, 2018MNRAS.473.5043E}.

Spectroscopic surveys have greatly enlarged the number of spectroscopic binaries by providing
millions of spectra. For~instance,     
\citet{2019AJ....158..155B} 
identified 3838 SB1 candidates using RAVE and Gaia DR2 data and~estimated 
the most probable orbital parameters of main-sequence dwarf candidates. 
\citet{2020ApJ...895....2P} 
discovered 19,635 APOGEE-SB1s and gave multiple possible
orbital solutions for each system. 
\citet{2020A&A...638A.145T} 
presented 12,760 binaries detected as SB2s in GALAH. 
\citet{2020A&A...635A.155M}
~found 641 FGK SB1 candidates at the 5$\sigma$ level in~GES.

LAMOST is a special quasi-meridian reflecting Schmidt telescope with an effective
aperture of 4\,m and a field of view of 20\,deg$^2$. The~unique design of LAMOST enables it
to acquire 4000 spectra in a single exposure \citep{2012RAA....12.1197C,2012RAA....12..723Z}.
Up until now it has obtained more than 11~million low-resolution ($R=1800$) spectra and more than 13 million
medium-resolution ($R=7500$) spectra. 

LAMOST is an invaluable provider of data on new spectral binary systems.
Numerous and consistently successful attempts have been made to detect unresolved binaries
in the LAMOST medium-resolution spectroscopic survey.
 Xiang et al. \citep{2021ApJS..253...22X} presented a data-driven method to estimate absolute magnitudes
for O- and B-type stars from the LAMOST spectra. The~inferred spectroscopic 
${K_{\rm s}}$-band magnitudes $M_{K{\rm{s}}}$
were  compared with the geometric $M_{K{\rm{s}}}$ magnitudes derived from Gaia 
parallaxes to identify binaries. Xiang et al.~\citep{2021ApJS..253...22X}  identified  1597 of the 16,002 LAMOST OB stars as binaries. 
\citet{2021ApJS..256...31L} examined radial velocities of over 1.3 million LAMOST DR7 medium-resolution blue-arm spectra and found 3133 spectroscopic binaries (SB) and 132 spectroscopic  candidate triples (ST). An overwhelming majority of these stars are new discoveries. Guo et al.~\citep{2022RAA....22b5009G} investigated the binarity of 9382~stars 
 measuring the equivalent widths of several spectral lines in the LAMOST medium-resolution spectra.
The binary fraction found by them ranges from  68 $\pm$ 8\,\% for O- to B4-type stars to 
44 $\pm$ 6\,\% for B8 to A stars. 
At the same time, the~results of searches for unresolved binaries in the low-resolution
LAMOST survey are rather scarce~\citep{2019RAA....19...64Q, 2023ApJS..269...41C}.

Thus, LAMOST is the largest spectroscopic survey in terms of the volume of data obtained
and the widest in terms of coverage of the sky.
The low-resolution LAMOST spectra collection is no less representative than the medium-resolution one.
There are not enough large spectroscopic surveys now to ignore this set of spectra. 
Additionally, LAMOST low-resolution spectroscopy has
accumulated data for nearly nine years; thus, there are multiple
observations for a significant number of sources. 
Examination of these sources may increase the number of known binary systems 
and give us additional information about the ensemble of binary stars in the Galaxy.
In the present study we aim to investigate the options to discover the population of unresolved binaries
using low-resolution spectra from LAMOST using synthetic~spectra.

If the stellar spectrum is the main information about the star, then, in~order to infer that the star is a binary, one needs to find the
deviation of its spectrum from that of a single star---an excess
due to the light from the second, fainter~component.

The methods of obtaining stellar parameters
from low-resolution stellar spectra were developed and widely applied to a substantial 
number of spectra in recent wide-scale spectroscopic surveys
\citep{2012MNRAS.426.2463L,2017ApJ...843...32T, 2024ApJS..272....2L}.

In parallel to these methods, other ones, based on multi-color
photometric data, were developed and widely used. The~reason for the
popularity of the latter, 
despite their accuracy in determining the physical parameters of stars not being high,
is that photometry,
as accurate as spectrophotometry, can be obtained with much smaller~telescopes.

The main trend in the development of stellar spectra analysis is
improvement in the accuracy of determination  of stellar physical
parameters (for example, it is possible to derive
the effective temperature of a star, $T_{\rm eff}$, with~an uncertainty
of several Kelvins). Therefore, the~methods based on the
analysis of the continuum are now out of use: they do not provide stellar parameters  with adequate precision.
However, in~this paper we will test the effectiveness of the method for detecting unresolved binary stars by comparing their observed continuum (in the LAMOST project) with synthetic spectra of single stars.

In Sections~\ref{sec:data} and~\ref{sec:data_cat}, we discuss
cross-identification of data of Gaia DR3 and LAMOST with selected
lists of close binaries and presumably single stars.
Section~\ref{sec:modeling} describes modeling of spectra of
unresolved binaries. Section~\ref{sec:results} discusses
results of testing approaches for distinguishing between single and
unresolved binaries, comparing LAMOST spectra
with sets of synthetic theoretical spectra of single stars, prepared in advance. 
Section~\ref{sec:6} presents our conclusions.

\section{Data: LAMOST~Spectra}
    \label{sec:data}

We use the LAMOST DR5  \citep{2019yCat.5164....0L} catalog of A-, F-, G- and K-type stars, 
which contains stellar parameters ($T_{\rm eff}$, $\log g$,
and $[\textrm{Fe/H}]$) for more than 5 million low-resolution ($R=1800$ at
5500\,{\AA}) stellar spectra. The~wavelength range of the spectra is about
3690--9100\,{\AA}. The~signal-to-noise ratio (SNR) is larger than 6 (in the $g$-band)
for the spectra obtained during dark nights, and~it is larger than 15 for the spectra
obtained  during bright  nights.
The acquisition of spectra and processing of LAMOST data 
are described in detail in a number of papers (see~\citet{Bai+2021} and references therein). 
We will not repeat them in this paper.

Currently, the~most frequently used methods 
for determining the parameters of stars
are those based on the analysis of the ratios of the depth of spectral line pairs 
(see, \mbox{e.g.,~\citet{2010ASPC..435..207P, 2019A&A...622A.114H, 2020ApJ...898...47F}).} 
The lines of the pair should have sufficiently close wavelengths in order to eliminate the effects of interstellar extinction to the greatest possible extent. 
Different line pairs are used for different spectral types. The~``classical'' line set is presented in~\citet{2000asqu.book.....C} (Table~15.3).

Line pairs analysis requires high-resolution spectra, but~this is not the case for LAMOST spectra.
\citet{2015RAA....15.1095L} developed a different method for them.
It is based on  minimizing the squared difference between the observational and
model spectra. This method is used both for the spectral
classification of stars in the catalog and for the determination of
the values of $T_{\rm eff}$, $\log g$, and~$[\textrm{Fe/H}]$.

It should be noted that while determining physical
parameters from stellar spectra we attempted to achieve the
best accuracy, but we can be satisfied with a rather rough parameter
determination, only to be able to detect excess radiation from the binary's second star. 
Thus, in~principle, all of the above methods are suitable for~us.

The quality of calibration of LAMOST spectra is important for solving our problem.
For the LAMOST flux calibration, the~LAMOST 2D pipeline picks out several high-S/N 
standards in the temperature range of 5750--7250 K and then obtains the spectrograph 
response curve (SRC) by comparing the observed spectra with synthetic spectra (using 
the corresponding parameters from 
the Castelli--Kurucz spectral library~\citep{2003IAUS..210P.A20C}). As~a result, 
the dereddening uncertainties of standards have an impact on the SRC derivation. 
This introduces uncertainties to all spectra of the spectrograph for one plate. 
For a few spectra ($\sim$2\%), flux calibrations are completed using the average 
response curve of each spectrograph (ASPSRC) method \citep{Du+2016}. 
This method has some uncertainties caused by variations in the shape 
of individual SRCs, which introduces uncertainties to the calibrated spectra. 
The statistical analysis presented in \citet{Du+2016} revealed that 
the total uncertainties caused by the LAMOST flux calibration 
in the spectra are less than 10\%.

The rest of the present paper deals with the method of detecting
unresolved binary stars in the LAMOST catalog by the analysis of
the spectral~continuum.

\section{Data: Known Single and Binary~Stars}
\label{sec:data_cat}

To investigate the possibility of distinguishing between single and
unresolved binary stars from LAMOST spectra, we
 need to know what might be systematic differences between actual LAMOST
 spectra  of single and binary stars. 
For this purpose, we compiled two data sets of LAMOST spectra. The~first one contains sources
 identified with known close binary stars, which cannot be resolved by LAMOST (the fiber diameter of
 LAMOST is about 3.3 arcsec). Another set contains presumably single stars identified by cross-matching the LAMOST DR5 source catalog with selected catalogs, as~described~below. 

\subsection{Reliable Unresolved Binary~Stars}

For the stars to be included in the test list of binaries for
modeling, we required them  to be definitely binary and to have known spectral types
of components. No multiple systems were accepted.
These requirements are met by the objects from the ninth catalog of Spectroscopic
Binary Orbits, SB9. 
\footnote{Pourbaix et~al.~\citep{2004A&A...424..727P} (\url{http://sb9.astro.ulb.ac.be}).} 
This catalog continues the series of compilations
of spectroscopic orbits carried out over  35 years by
A. Batten and his collaborators and is continuously updated.
As of 2023, the~catalog contains about 5000 orbits for about 4000~systems. 
Note that SB9 binary stars usually contain two similar components, i.e.,~with mass ratio $m_2/m_1 \simeq 1$.
There are  10 stars common to the LAMOST DR5 catalog. Two
observations were found in the catalog for one of these stars;
thus, in~total, we possess 11~spectra.

\subsection{Reliable Single~Stars}
\label{sec:singles}

We selected from several catalogs single stars that were highly reliably not 
binary or multiple. These catalogs are listed below with brief~comments:

\begin{itemize}

\item  Compilation of spectra of single stars, mostly Praesepe and M67 spectral standards
\citep{1995AJ....109.1379A}. There is $N_{ov}$ = 1 star overlapping with the LAMOST DR5~catalog.

\item ELODIE.3.1---an updated release of the library published
in~\citep{2007astro.ph..3658P}. \mbox{$N_{ov}$ = 15}.

\item The Indo-US Library of Coud\'e Feed Stellar
Spectra~\citep{2004ApJS..152..251V} containing spectra of 1273~stars obtained using the 0.9 m Coud\'e feed telescope at Kitt
Peak National Observatory. \mbox{$N_{ov}$ = 30}.

\item LCO-SL, the Las Campanas Observatory Stellar
Library\footnote{\url{https://sl.voxastro.org/}}, which is the most comprehensive
near-infrared spectral library with 1300+ spectra obtained between 2013 and 2017
using the Folded-port InfraRed Echellette (FIRE) spectrograph
operated using the 6.5 m Magellan Baade telescope. $N_{ov}$ = 12. 

\item MILES---the library of spectra obtained by the 2.5 m Isaac Newton
Telescope used, in~particular, for~stellar population synthesis
\citep{2011A&A...532A..95F}. $N_{ov}$ = 25.

\item Hubble's Next-Generation Spectral Library (NGSL) containing
$\sim$1000 spectra of 380~stars of assorted temperature,
gravity, and~metallicity. Each spectrum covers the wavelength
range 0.18--1.03 $\upmu$~\citep{2007ASPC..374..409H}.
\footnote{\url{http://archive.stsci.edu/prepds/stisngsl/} (accessed on 28.07.2025).}
$N_{ov}$ = 6.

\item The Tian Shan (Alma-Ata) Sternberg Astronomical Institute
photometric $WBVR$ catalog of 13,586  bright northern-sky
stars~\citep{1996BaltA...5..379K}.
The catalog contains both  single stars and double (multiple) systems. 
For single stars, $N_{ov} = 102$.

\item The latest issue of the astrometric fundamental star catalog,
FK6~\citep{1999VeARI..35....1W}, a~combination of results from
ground-based observations and from the Hipparcos project of space
astrometry. $N_{ov} = 1$.

\end{itemize}

After exclusion of overlapping objects, the list contains 166 stars.
For these stars, we find 278 observations (spectra) in the
LAMOST DR5 catalog. Hereafter, we will refer to this sample as the main list of single stars, Main\_Single.

\subsection{Presumably Single and Unresolved Binary~Stars}
\label{sec:Gaia}

The number of reliable single stars identified with LAMOST DR5 objects was too low to provide a representative sample, so we added a number of stars that we presumed to be single 
and to be unresolved binaries, employing the data from Gaia~DR3. 

\subsubsection{Presumably Single~Stars}
\label{sec:Gaia-single}

For this purpose, we cross-matched the LAMOST DR5 A-, F-, G-, and K-type star catalog \cite{2019yCat.5164....0L} with Gaia DR3, using a standard X-Match  CDS procedure with a radius of matching $1^{\prime\prime}$, under the condition that there was no other Gaia source within the radius of 
$4^{\prime\prime}$ from the LAMOST star, to~avoid doubtful matches. We selected only Gaia DR3 sources with five-parametric solutions without ``duplicate source'' and ``variable source'' flags, which, by~Gaia DR3 classification, have zero probability of being a galaxy or a quasar 
($classprob\_dsc\_combmod\_quasar=0$
and $ classprob\_dsc\_combmod\_galaxy=0$) and~are most probably single stars ($classprob\_dsc\_combmod\_star>0.99$) \citep{2023A&A...674A..28F}. 
Further, we used photometry data to filter out probable mismatches based on comparison of photometric magnitude measurements. This filter is tentative, as~the data on LAMOST sources after automatic processing of their spectra 
may contain several photometric magnitudes in different bands and not all of them are obligatorily presented.
So, we filtered out stars with one of the LAMOST stellar magnitudes differing from Gaia G-magnitude by 
more than 3 rms in the entire sample of matched LAMOST--Gaia~stars.

Next, we have filtered out the stars matching with Gaia sources with any indication of being non-single, astrometrically, photometrically, or~spectroscopically. We have selected stars in close vicinity of the Sun (having parallax $\varpi \ge 5$, e.g.,~approximately within 200 pc)  to make possibly unresolved components discoverable. It was required that the matched sources in Gaia had 
Renormalized Unit Weight Error $ruwe<1.4$ \citep{2021A&A...649A...2L} and~corrected flux excess factors indicating no inconsistency between BP, RP, and G fluxes \citep{2021A&A...649A...3R}, since it was demonstrated that these factors may indicate duplicity of the source (see, e.g., \citet{2020MNRAS.496.1922B}). Finally, we have checked that the remaining dataset did not contain stars marked as Gaia unresolved non-single sources \citep{2023A&A...674A..34G}.

The resulting list of presumably single Gaia sources cross-identified with LAMOST contains 6220 stars. 
Hereafter, we will name this sample Gaia list of single stars Gaia\_Single. 

\subsection{Gaia Non-Single~Stars}
\label{sec:Gaia-binary}
 
To increase the number of unresolved binary stars identified with LAMOST DR5 objects, we added a number of stars that are marked as presumably unresolved binary stars in the Gaia DR3 non-single star (NSS) catalogs~\citep{2023A&A...674A..34G}.  These catalogs include sources revealing signs of periodical changes in astrometric, photometric, and/or spectroscopic
Gaia data, which are interpreted as their binarity or~multiplicity. 
 
For this purpose, we cross-matched the LAMOST DR5 A-, F-, G-, and K-type star catalog~\citep{2019yCat.5164....0L} with Gaia DR3 NSS sources, using a standard X-Match  CDS procedure with a radius of matching of $1''$ on the condition that there was no other Gaia source within a radius of $4''$ from the LAMOST star, to~avoid doubtful matches. Further, we used photometry data to filter out probable mismatches based on comparison of photometric magnitude measurements. This filter is tentative, as~the data on LAMOST sources after automatic processing of their spectra 
may contain several photometric magnitudes in different bands and not all of them are obligatorily presented. So, we filtered out stars with some of the LAMOST stellar magnitudes differing from Gaia G-magnitudes by more than 3 rms in the 
entire sample of matched LAMOST--Gaia stars. 
Next, we selected the stars matching with Gaia sources in close vicinity of the Sun (having parallax 
$\varpi \ge 5$, e.g.,~approximately within 200 pc). 
 
The resulting list of presumably unresolved Gaia sources cross-identified with LAMOST contains 2278 stars. 

\section{The Method for Comparing Model Spectra with the LAMOST Spectrum in the Terms of Their~Continua}
\label{sec:modeling}
\unskip

\subsection{Description of the Method in~General}

Among the astronomical software, there are a number of well-established programs 
for working with spectra, such as ULySS~\citep{ULySS-1,ULySS-2}, SME~\citep{SME-1,SME-2,SME-3}, 
FERRE~\citep{FERRE}, The~Payne~\citep{Payne}, and RVSpecFit~\citep{RVSpecFit}. 
It might be possible to solve our problem using the FERRE package, 
which is unnecessary for our purposes. So, we applied our own code in Python v.~3.11.

LAMOST spectra and synthetic spectra are defined for different
wavelength sets and normalized differently. In~order to compare spectra, 
the former should be recomputed for the same set of wavelengths and
presented with the same normalization. At~this stage, the~defects that are
sometimes present at the edges of the LAMOST spectra can be detected and
removed.  

The criterion for two spectra to be considered similar is their mutual rms
deviation.
Among the synthetic spectra, we search for the one closest to the studied LAMOST spectrum according
to the above criterion, with~an account of
interstellar reddening, within~the limits acceptable for the star
under study. We call such a reddened synthetic spectrum  ``optimal''.

If divergence of the observed LAMOST spectrum and the optimal
one is not large,  the star under study is either a single star
or a binary with a faint component that cannot be
distinguished at the background of the companion. 
This category
also includes the case of double stars with almost identical
components whose spectra differ only slightly. 
Such a star will have a luminosity exceeding that of the more massive component of the binary,
but the luminosity increase will not be recognized by us.
If the spectra differ strongly, the LAMOST star can be peculiar (single or
binary). 
A~similar effect can be also caused by a strong defect in the LAMOST~spectrum.

For an unresolved binary, the~variance of the observed spectrum
and the closest synthetic spectrum can look different in the 
following two cases.
If one of the components is considerably brighter than the second one, the~synthetic spectrum will be 
close to that of the bright component.
An increase in the brightness difference of components hampers the discovery of unresolved~binarity.

The second case corresponds to the occurrence when the
radiation fluxes from components of an unresolved
binary in the LAMOST working range are approximately the same,
while the  spectral types of them differ considerably. 
The synthetic spectrum consists of a component with one maximum, while  
the binary spectrum is the sum of two such components. 
The difference of observed and synthetic
spectra will have a ``wave-like'' shape.
This case is much more favorable for the detection of~binarity.
\begin{figure}[H] 
     \includegraphics[width=0.7\textwidth]{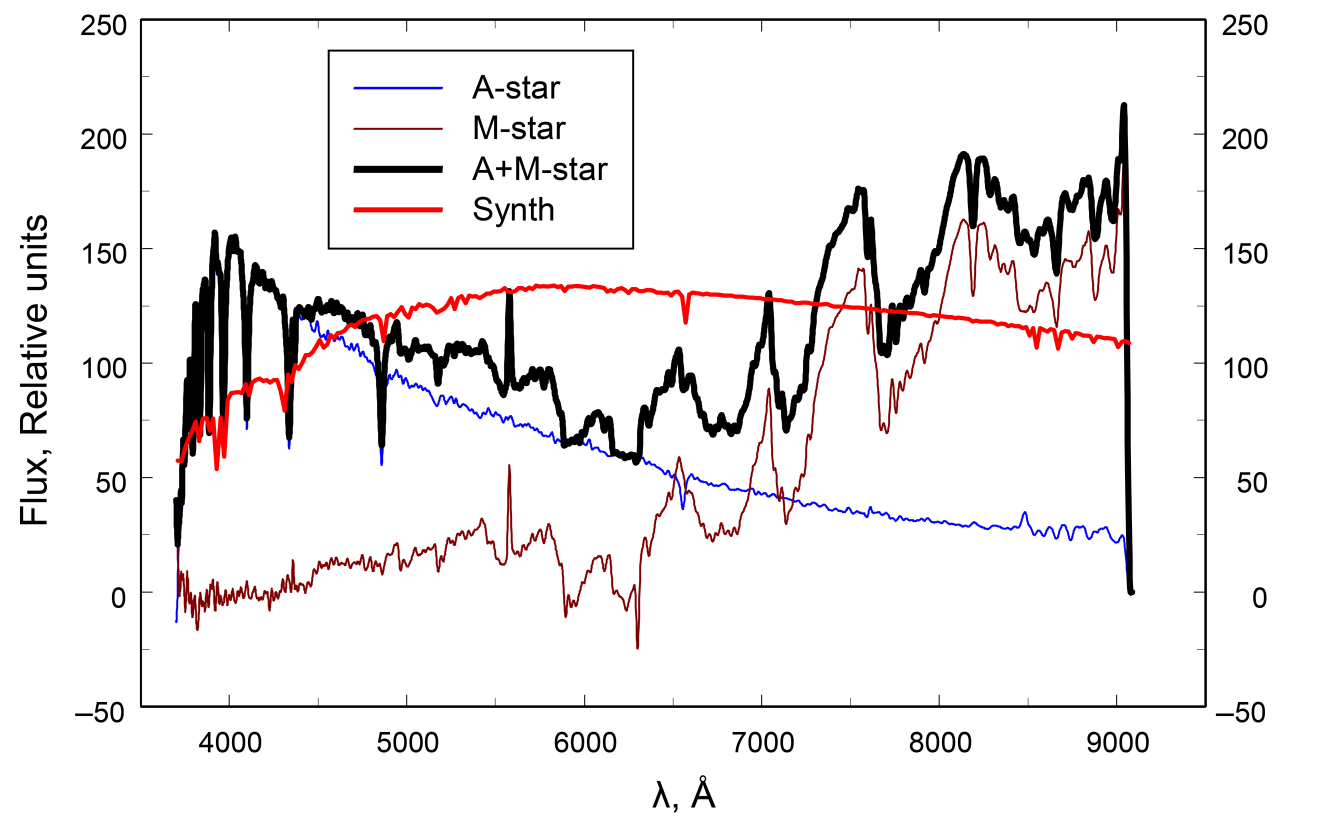} 
     \includegraphics[width=0.7\textwidth]{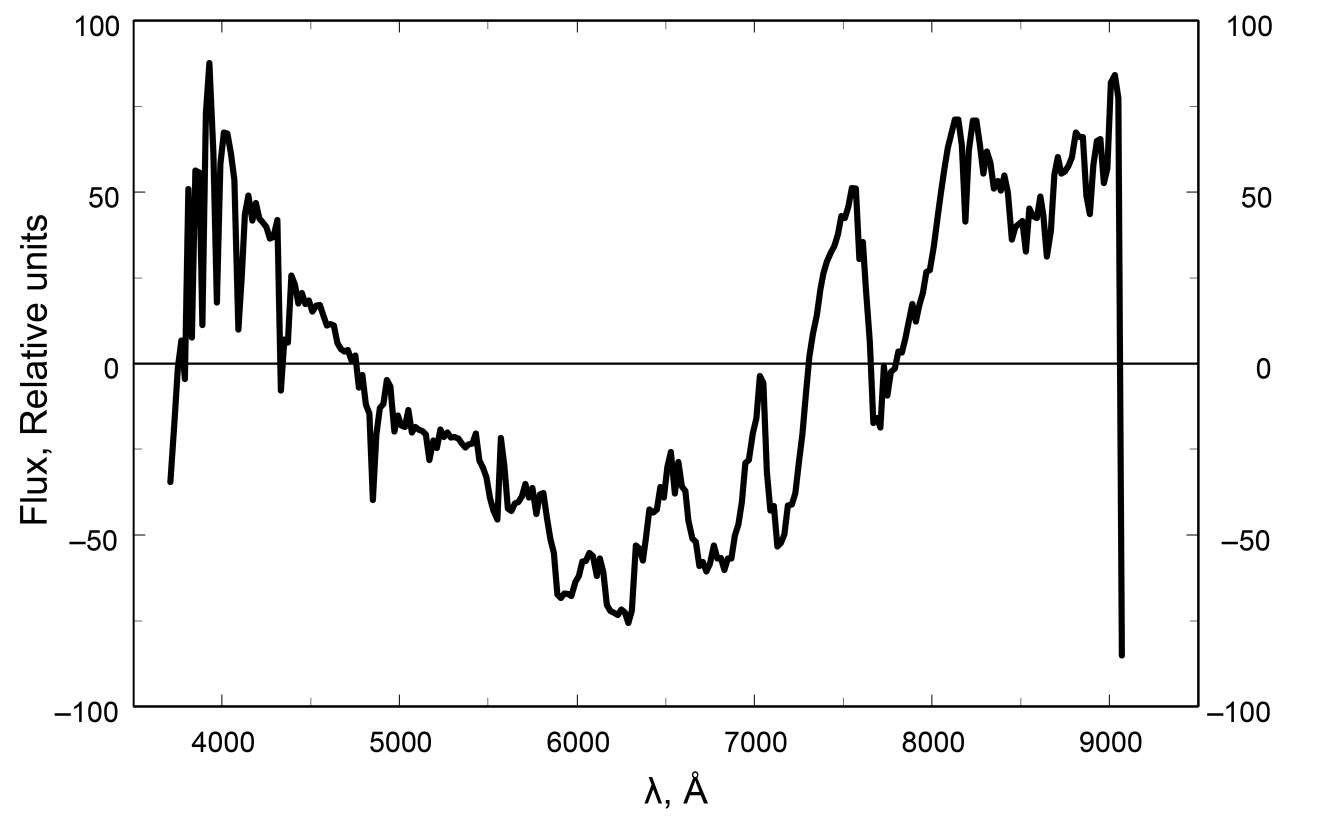} 
\caption{Upper 
 panel---the combination of spectra of stars belonging to A1V and M6
spectral types from the LAMOST catalog as a spectrum of an
unresolved binary with equal radiation fluxes from both components 
and the closest synthetic spectrum from the Castelli--Kurucz library (CKL).
 Lower panel---the difference between the spectrum of the unresolved
binary with A1V and M6 components shown in the upper panel and
the closest synthetic spectrum from CKL.}
\label{fig:A+M}
\end{figure}
\unskip
Let us check if the suggested method is productive in
the clearly expressed latter case. Simultaneously, we will be
able to evaluate the upper limit for the mutual deviations of
spectra between the binary and optimal single synthetic stars. 
For this purpose, we selected real A1V- and M6-type stars from the LAMOST DR5 catalog.
Such stars are the most different in the catalog.
For more information on them, see Table~\ref{tab:A+M}.
The maximum of the continuum in the spectrum of the A1V star is located 
in the short-wavelength region, while in the spectrum of the M6 star it is at
long wavelengths (see Figure~\ref{fig:A+M}).
The sum of the spectra of such a pair will differ from the spectrum of a star of any
spectral type. However, the~latter difference will be not large, if~the radiation flux of one of
the stars is much lower than that from the companion. 
The~difference is at the maximum,
if the fluxes are equal. In~order to illuminate the latter case, the~fluxes were ``equalized''
by multiplication of one of the spectra by an appropriate factor.
The spectra of both stars and the combined spectrum
of the artificial binary are shown in the upper panel of Figure~\ref{fig:A+M}.
The combined spectrum has two maxima situated at the edges of the LAMOST
spectrograph's working~range.

\begin{table}[H]
\centering
\caption{Stars from the LAMOST DR5~catalog.
\label{tab:A+M}}
\begin{tabular}{ccccc}
\hline\hline
\textbf{Obs\_ID} & \textbf{Designation} & ${\mathbf \alpha}$ & ${\mathbf \delta}$ & \textbf{Sp.} \\
\hline
404058 & J004541.11 + 405100.1 & 11.421 & 40.850 & A1V \\
113226 & J221422.57 + 005400.8 & 333.594 & 0.900 & M6 \\
\hline\hline
\end{tabular}
\end{table}

We searched for the optimal synthetic spectrum for all possible
combinations of the parameters $T_{\rm eff}$, $\log g$, and~$[\textrm{Fe/H}]$ available in the New Grids library of ATLAS9
Model Atmospheres~\citep{2003IAUS..210P.A20C}, hereafter the
Castelli--Kurucz library (CKL); see Section~\ref{sec:CKL} below.
Interstellar extinction was incrementally increased by $\Delta A_v=0.1$
from  0 to 1. The~spectrum with the following parameters was found optimal:
\begin{center}
$T_{\rm eff} = 5500$\,K, 
$\log g=2.5$, $[\textrm{Fe/H}]=-1.5$, $A_v=1.0$.
\end{center}

\noindent
It is also shown in the upper panel of Figure~\ref{fig:A+M}. Naturally, this
spectrum has a single maximum located between the maxima of the ``binary's'' 
combined spectrum, near~the middle of the working range of the LAMOST~survey.

The difference between the combined spectrum of the ``binary'' and the
optimal synthetic spectrum (lower panel of Figure~\ref{fig:A+M}) has a ``wave-like'' shape with a rather
large amplitude: the rms deviation for a point of the synthetic
spectrum is as large as 37\% of the mean signal level. This value,
close to rms, may be accepted
as an estimate of the upper limit for the possible
deviation between the spectra in our approach. Figure~\ref{fig:A+M} suggests 
that, in~principle, the~applied method may be used to search for stellar~duplicity.

To establish a threshold between low and high levels of
deviations between spectra, a~statistical analysis is needed. For~this purpose, we selected two special star samples from the
LAMOST catalog, so it was known for the first of them from
\textit{a priori} information that the stars were single and~for the
second of them that the stars were~binary.

\subsection{The Library of Synthetic~Spectra}
\label{sec:CKL}

The spectra in the CKL were computed for 1221 wavelengths between 9.09 nm and 160~$\upmu$m. The~main parameters of stellar atmospheres in the library vary within
the following limits:
\begin{itemize}
\item $T_{\rm eff}$ takes 76 values from 3500~K to 50,000~K. The~increment in $T_{\rm eff}$ is not uniform: it increases from  $\Delta T=250$~K for low temperatures
to $\Delta T=1000$~K for high temperatures;

\item $\log g$ takes 11~values, from~0 to 5, with~the uniform increment $\Delta\log g=0.5$;

\item Not every
combination of these two parameters is possible; the~allowed
combinations form a triangle in the $T_{\rm eff}$---$\log g$
plane. See \citet{2003IAUS..210P.A20C} for details;

\item $[\textrm{Fe/H}]$ takes values from $-1.5$ to $+0.5$, with~a
uniform increment $\Delta[\textrm{Fe/H}]=0.5$. 

\end{itemize}

We use only a part of a particular synthetic spectrum within~the LAMOST sensitivity
range 3700 \AA\ $< \lambda <$ 9097 \AA. This interval contains 270
points of the synthetic spectra from the CKL library with
wavelengths 3710~\AA\ $\aplt \lambda \aplt 9090\ $\AA.

\subsection{Removal of the Outliers at the Edges of LAMOST~Spectra}

Each LAMOST spectrum contains 3908 data points in the
3699.9863 \AA~$ < \lambda < 9097.04 $~\AA\ range. We use complete~spectra.

\begin{figure}[h] 
  \hspace{-6mm} \includegraphics[width=0.9\textwidth]{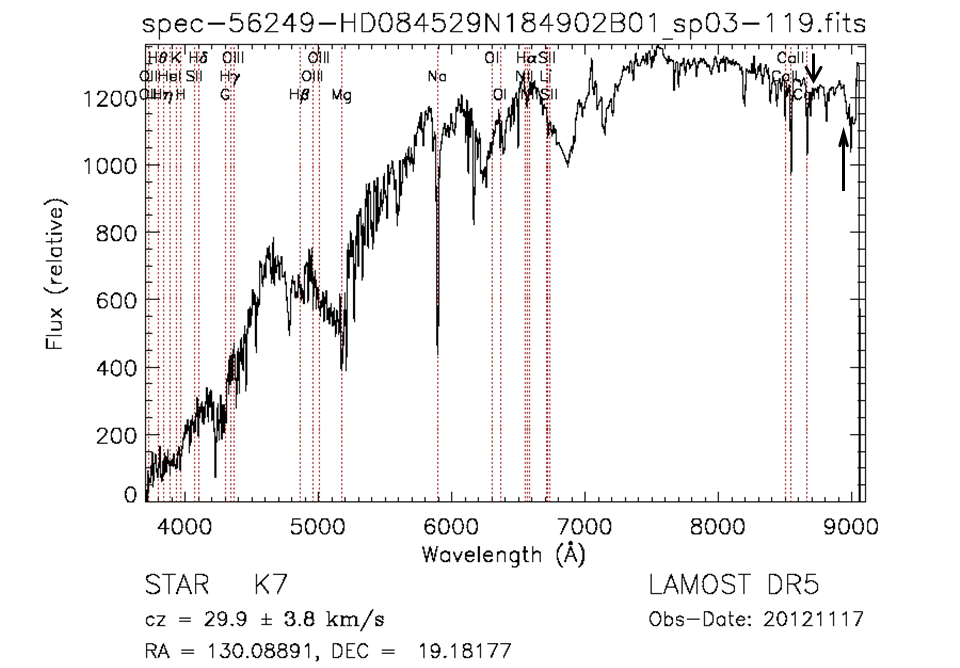} 
    \caption{A LAMOST
 spectrum with outliers. Vertical dotted lines in this and
		the next figure mark the wavelengths of several spectral lines.
		The down arrow indicates the short-wave boundary of the region 
		at the edge of the spectrum in which outliers are sought. 
		The up arrow is the short-wave edge of the outlier in the presented spectrum.
		\label{fig:right}}
\end{figure}

\begin{figure}[h] 
  \hspace{-1mm} \includegraphics[width=0.9\textwidth]{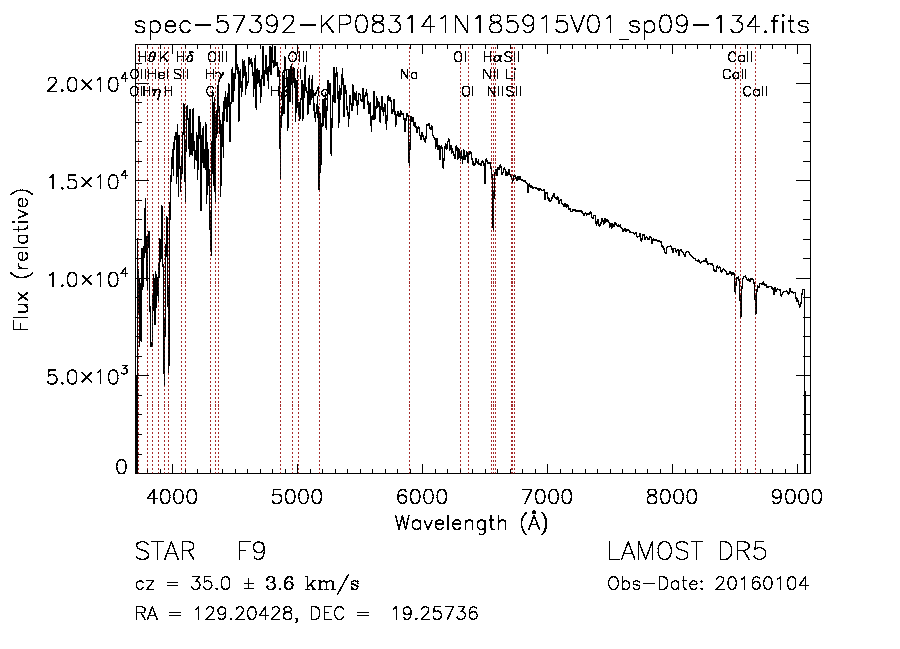} 
    \caption{A LAMOST  spectrum without~outliers.\label{fig:LAMOST}}
\end{figure}

Rather frequent defects of the LAMOST spectra are outliers at the
long-wavelength edge of the spectrum. 
Among 
several hundred randomly selected spectra from the LAMOST archive that we viewed, outliers were present in $\sim$25--30\% of the images. Synthetic spectra of stars with similar parameters in this region do not show any peculiarities in the behavior of the continuum. 
These outliers can be positive or negative and vary in width.
An example of a spectrum with outliers is displayed in Figure~\ref{fig:right}, while one without outliers is shown in  
Figure~\ref{fig:LAMOST}\footnote{Both spectra are taken from the LAMOST DR5 site.}.
The reason for the occurrence of these emissions is unknown.

We detect the presence of an outlier in the following~way:
\begin{itemize}
\item 
We check whether there is an outlier in the $\lambda > \lambda_{\textrm{right}}=8700$~\AA\ range. 
This value is chosen based on the results of visual analysis of the spectra.

\item 
We search a data point $\lambda_R$ in the spectrum, closest to 
$\lambda_{\textrm{right}}$ at longer $\lambda$. 
For most LAMOST spectra, the~corresponding wavelength is $\lambda_R = 8701.62$~\AA.
The flux at this $\lambda$ is $F_R = F(\lambda_R)$.

\item 
If at any point of the LAMOST spectrum with $\lambda_i > \lambda_R$, the~deviation of the flux from $F_R$
exceeds the pre-selected value, $\varepsilon$, i.e.,
mod$(F(\lambda_i)/F_R-1)) > \varepsilon$,
we assume that an outlier is present. In~this study, we accept  $\varepsilon=0.2$. 

\end{itemize}

If an outlier is detected, we reject the part of the spectrum with $\lambda > \lambda_R$,
since there are no strong stellar lines in the 8700~\AA\ $< \lambda <$ 9100~\AA\ range of stellar~spectra.

\subsection{Interpolation and Normalization of the~Spectra}

LAMOST spectra and synthetic spectra in the CKL are defined for different wavelength sets.
In addition, the~set of $\lambda$s  in the LAMOST spectra is more than ten times
larger than in the CKL synthetic spectra. 
Therefore, we interpolate the latter to the LAMOST set of wavelengths.
Next, we smooth the LAMOST spectra in order to reduce the 
discrepancies in spectral comparisons.
We use smoothing because the LAMOST spectra are quite noisy 
(see Figures~\ref{fig:right} and~\ref{fig:LAMOST}).
We apply a moving average using a triangular profile of weights according to the formula

\begin{equation}
	\left<F_i\right> =
	\frac{\sum\limits_{j=i-s}^{i+s} \left(1-\frac{|i-j|}{s}\right)F_j}{2s+1}.
\end{equation}
Here,  $i$, $j$~are the numbers of data points in the spectrum; $s$~is
the smoothing width, taken as a parameter. We compute the
smoothed spectrum for the same wavelengths as those in the initial
LAMOST spectrum.
A side effect of this procedure is removal of narrow and
not very intense~lines from the LAMOST spectra. 

At the edges of the LAMOST spectra, at~$i<i_{\min}^{\mathrm{(LAMOST)}}+s$ or
$i>i_{\max}^{\mathrm{(LAMOST)}}-s$, no smoothing was~applied.

We tried the smoothing parameters $s=0$ (no smoothing), 
$s=3$, and $s=10$ and found that without smoothing the discrepancy 
in the comparison of spectra is noticeably larger, while with 
$s=3$ and $s=10$ the results are comparable. 
Further calculations were carried out with $s=3$.

At the next stage, we interpolated the smoothed spectra to the set
of wavelengths of the synthetic spectra  within the
LAMOST sensitivity range. Piecewise linear interpolation was used.
We designate fluxes in the interpolated spectrum as $\tilde
F_{i^{\mathrm{(CKL)}}}$.

When comparing the spectra, we had to not only reduce them to a
common set of wavelengths but~also perform their normalization.
We left the fluxes in the interpolated  LAMOST spectra
$F_i^{\mathrm{(CKL)}}$ without changes.
The values of fluxes
in the synthetic spectrum were multiplied by the coefficient:

\begin{equation}
	C = \frac{
	  \sum_{ j=i_{\min}^\mathrm{(CKL)} }^{ i_{\max}^{\mathrm{(CKL)}}-1 }
		  F_j^{\mathrm{(CKL)}}
	}{
	  \sum_{ j=i_{\min}^\mathrm{(CKL)} }^{ i_{\max}^{\mathrm{(CKL)}}-1 }
		  \tilde F_j^{\mathrm{(CKL)}}
	}.
\end{equation}

\subsection{Four-Parameter Representation of Stellar Synthetic Spectra Taking into Account Interstellar~Reddening}
\label{sec:4D-set}

We did not vary additional parameters that influenced appearance of the spectra 
less than $T_{eff}$, $\log g$, and [Fe/H]. 
The ranges and increments of these parameters
in the CKL library  are described in Section~\ref{sec:CKL}.
For the value of turbulent velocity in the photosphere, we assumed
$v_{\rm turb}$=2.0~km/s. 
Stellar axial rotation was not taken into~account.

The observed spectrum of a star can differ from the synthetic one
because of interstellar extinction; its amount, described by the parameter $A_V$, is the fourth parameter in the total
representation of stellar~spectra.

To take into account interstellar reddening, we
used the law suggested by \citet{1989ApJ...345..245C}.
In the LAMOST sensitivity range, it is described by the
following relations:
\begin{equation}
	\begin{array}{l}
		x=1/\lambda_\mu,\\
 		y = - 1.82,\\
 		a(x) = 1 +0.17699y -0.50447y^2 -0.02427y^3 +0.72085y^4 \\ \qquad\qquad +0.01979y^5 -0.77530y^6 +0.32999y^7, \\
		b(x) = 1.41338y +2.28305y^2 +1.07233y^3 -5.38434y^4 \\ \qquad\qquad -0.62251y^5 +5.30260y^6 -2.09002y^7, \\

		\dfrac{A(\lambda)}{A_V} = a(x) + \dfrac{b(x)}{R_V}, \\
		R_V \equiv \dfrac{A_V}{E_{B-V}}.
	\end{array}
	\label{eq:Extinct_Law}
\end{equation}

We applied the standard assumption that the ratio $R_V$ was
constant, namely, $R_V = 3.1$. For~the wavelength corresponding to
the $V$ band, we assumed $\lambda_V = 550$~nm.
The four-parameter (4D) set of spectra was computed for all spectra
of the CKL library in the following~way:

\begin{enumerate}

\item 
For every star, we subdivided  the range of possible values of
interstellar absorption $[A_{V,\min}, A_{V,\max}]$
into $n(A_V)=20$ intervals, and for every  input synthetic spectrum from the CKL
library we computed 21 spectra with different values of $A_V$. 

\item 
Next, we formed the relative absorption line, i.e.,~for all
270 wavelengths of CKL spectra that were within the LAMOST range
(see Section~\ref{sec:CKL}) we computed $A(\lambda)$ using
Equation~(\ref{eq:Extinct_Law}) with $A_V = 1$.

\item 
For a particular star, we found the acceptable range of
interstellar absorption, $[A_{V,\min}, A_{V,\max}]$. The~edges  of
this interval depended on stellar coordinates $\alpha$,~$\delta$, which were well known and
presented in the LAMOST catalog, and on the distance to the star $d$ absent in the~catalog.

To determine the distance $d$, we used several~approaches:
\begin{enumerate}
\item 
The majority of stars from the LAMOST catalog could be identified with the stars from
the  Gaia DR3 catalog~\citep{2021A&A...649A...1G, 2023A&A...674A...1G} that currently contains the
most accurate stellar parallaxes $\varpi^{(G)}$ and their uncertainties $\sigma_\varpi^{(G)}$.
Using them, we computed possible limits for the distance to the star:
\begin{equation}
d_{\min} = \dfrac{1}{\varpi^{(G)} + 3\sigma_\varpi^{(G)}}, 
\qquad
d_{\max} = \dfrac{1}{\varpi^{(G)} - 3\sigma_\varpi^{(G)}}.
\end{equation}
If the computed $d_{\min}$ turned out to be negative, we set $d_{\min}=0$.

\item 
If a LAMOST star had no identification in the Gaia DR3 catalog, then, for~some single
stars, we took information on the distance or interstellar absorption directly from the original catalogs (see Section~\ref{sec:singles}).

\item 
If identification of a LAMOST star with the Gaia DR3
catalog and  any information on the distance to
it were lacking, we assumed $d_{\min} = 0,  d_{\max} = 5\mathrm{~Kpc.}$
\end{enumerate}

\item
The values of $d_{\min}$ and $d_{\max}$ together with the stellar
coordinates $\alpha$ and $\delta$ were uploaded to
the site 
\url{http://stilism.obspm.fr} (accessed on 05.12.2024),  where an interactive procedure  permitting us to determine the interstellar reddening
$E_{B-V}$ from a 3D map of interstellar absorption described
in~\citet{2014A&A...561A..91L, 2017A&A...606A..65C,
2018A&A...616A.132L} is possible.
This procedure returns the limits of interstellar reddening,
$E_{B-V,\min}$, $E_{B-V,\max}$, possible for the star, with
corresponding distances $d_{\min}$ and $d_{\max}$ and the
limiting distance $d_{stil}$ covered  by the interactive
interstellar absorption map on the site. 
When $ d_{\max} > d_{stil}$, we assumed $ d_{\max}= 2 \times d_{stil}$,
having in mind  that factor 2 results in the maximum possible absorption for barometric
dependence of the latter on the $z$-coordinate.

\item 
We recomputed the derived limits of the interstellar
reddening $E_{B-V}$ into the limits of the interstellar absorption
$A_{V,\min} = R_V E_{B-V,\min}$ and
$A_{V,\max} = R_V E_{B-V,\max}$.

\item 
For every spectrum from the CKL library defined  by the
parameters $T_{\rm eff}$, $\log g$, and~$[\textrm{Fe/H}]$, and~for
all wavelengths in the sensitivity range of the LAMOST survey, we
computed reddened spectra for the following values of interstellar absorption:

\begin{equation}
	A_{V,i} = A_{V,\min} + i \frac{ A_{V,\max} - A_{V,\min}}{n(A_V)}\,,
	\label{eq:Avi}
\end{equation}
where the index $i$ takes the values $i=0\dots n(A_V)$ and
where $n(A_V)=20$ is the number of intervals into which the range of interstellar
absorption is split.
To compute
the intensity of light for the reddened synthetic spectrum, we used the
\citet{1856MNRAS..17...12P} formula
$I(A_{V,i},\lambda) = 10^{ -0.4 A(\lambda) A_{V,i} }
i_{CKL}(\lambda)$, where $i_{CKL}(\lambda)$~are intensities of
stellar light given in the CKL library without account of the interstellar absorption.
\end{enumerate}

Above, we described the part of the procedure we developed where we
estimated the limits of interstellar absorption for a particular star
using data not contained in the LAMOST catalog~itself. 

Certainly, uncertainties and systematic trends in the LAMOST 
flux calibration may also affect the results we obtain. Taking them into account is very important. 
However, we do not have the necessary initial data to study this influence. 
We assumed that the standard calibrations of the spectra in the LAMOST project 
were carried out with the highest quality.

\subsection{The Search for Optimal Synthetic Spectrum in the Four-Parameter Set of
 the Reddened Synthetic~Spectra}

The procedure of the search for the optimal synthetic spectrum in
the 4-parameter set of the reddened synthetic spectra has two~steps.

First, in~the synthetic set  generated for a particular LAMOST star (see Section~\ref{sec:4D-set}), we search for the spectrum
with minimum deviation from the observed spectrum.
Second, we attempt to improve parameters of the synthetic
spectrum by  interpolation among the spectra of the
4-parameter set with the most similar parameters.
Let us consider these steps one by~one.

\subsubsection{The Search for a Spectrum with the Lowest Rms Deviations in the 4-Parameter Set of Synthetic~Spectra}

We searched for the spectrum with lowest rms deviations from the
observed LAMOST spectrum in the 4-parameter set of reddened
synthetic spectra using two different techniques described below,
along with their advantages and~drawbacks.

First, we applied the discrete descent method.
Implementation of this method has \mbox{several~steps}:

\begin{enumerate}
\item 
For a selected LAMOST spectrum, we found in the CKL library a spectrum with the
parameters $T_{\rm eff,\,0}$, $\log g_0$, and~$[\textrm{Fe/H}]_0$,
closest to the catalog parameters $T_{\rm
eff}^{(\mathrm{LAMOST})}$, $\log g^{(\mathrm{LAMOST})}$, and~$[\textrm{Fe/H}]^{(\mathrm{LAMOST})}$. For~the initial value of
interstellar absorption $A_{V,\,0}$, we selected the value from
the sequence of acceptable values provided by Formula
(\ref{eq:Avi}) that was either the closest to the interstellar
absorption corresponding to the star's mean parallax $\varpi$ (if
the star was identified with the Gaia catalog) or the closest to
the mean interstellar absorption known from the other sources (see
Section~\ref{sec:4D-set}).

\item 
The so-called ``current spectrum'', containing intermediate
optimization results, is introduced. We designate parameters of
the current spectrum as $T_{{\rm eff},\,i}$, $\log g_i$,
$[\textrm{Fe/H}]_i$, and~$A_{V,\,i}$ for the spectrum used at the first step of the
optimization procedure $i=0$.

\item 
\label{enum:begin_loop} 
The deviation $\sigma_C$ is computed
for the current spectrum: $\sigma_C = \sigma(T_{\rm eff,\,i}, \log
g_i, [\mathrm{Fe/H}]_i, A_{V,\,i})$.

\item 
We compute deviations for eight adjacent spectra,
$\sigma_{near8}$, that differ from the current spectrum by their
value of one of the parameters in the 4-parameter set of synthetic
spectra. For~example, $\sigma_{T+} = \sigma(T_{{\rm eff},\,i+1},
\log g_i, [\mathrm{Fe/H}]_i, A_{V,\,i})$~is the deviation for the
spectrum with $\log g_i$, $[\mathrm{Fe/H}]_i$, and
$A_{V,\,i}$ similar to those of the current spectrum, but~$T_{\rm eff}$ is the closest one, exceeding $T_{{\rm eff},\,i}$.

If the parameters of the current
spectrum are at the edge of the 4D set, i.e.,~are at the maximum or
minimum possible level, then~the current spectrum will have only
one neighbor in this~direction.

\item 
If the deviation for one of these neighbor spectra is lower
than that for the current spectrum,
$\min(\sigma_{near8}) < \sigma_C$, 
then the spectrum with the lowest deviation is chosen
as the current spectrum, and~the minimization procedure returns
to step~3.

\item 
If $\sigma_C \le \min(\sigma_{near8})$, we additionally
calculate deviations for 72 adjacent spectra, $\sigma_{near72}$,
so that these spectra simultaneously have from 2 to 4 parameters
differing from the current spectrum by one step. Similarly to
calculating deviations $\sigma_{near8}$, if~the current point is
at the edge of the 4D distribution of the parameters, it will have
less than 72 additional neighbors.
If the deviation for one of adjacent spectra of the
additional set is lower than that for the current spectrum,
$\min(\sigma_{near72}) < \sigma_C$, then we select the spectrum
with the lowest deviation as the current spectrum. The~minimization procedure returns to step~3. 

\item 
If $\sigma_C \le \min(\sigma_{near72})$, the~procedure of
the search for the spectrum with the minimum rms deviation in the
4-parameter set is completed, and the desired  spectrum $F^{(opt)}$
is contained in the current spectrum.
\end{enumerate}

Second, we applied the method of exhaustive search, by~which
we computed deviations for all reddened spectra of
the 4-parameter set and selected the  one with the lowest~deviation.

\subsubsection{Comparison of the~Methods}

The discrete descent method is faster than the method of
exhaustive search by almost three orders of magnitude. Performing
the exhaustive search is, from~a computational point of view,
the most effort-consuming part of the technique we~applied.

The exhaustive search is guaranteed to find the global extremum of
the multi-dimensional deviation function $\sigma(T_{\rm eff}, \log
g, [\mathrm{Fe/H}], A_{V})$ at a discreet and limited parameter
grid, no matter how many local extrema there~are.

The discrete descent method is able to find the global extremum in two
cases. First, if~the deviation function is monotonous, i.e.,~if it possesses a
single extremum in its domain, the~result of optimization will not
depend on the choice of the initial spectrum. Second, if~the deviation function possesses several local minima and one of them, namely, the~deepest one, is the global minimum, we will
find the global extremum only if the parameters of the starting
spectrum happen to be on the slopes of the global~extremum.

It turns out that the methods of discrete descent and exhaustive
search give different results in approximately 10 to 15\%
of all cases, when the deviation function has several extrema and the
starting spectra are different.
Unfortunately, we had to reject the discrete descent method and
apply the much more effort-taking, but~guaranteed successful,
exhaustive~search.

\subsection{Improving the Optimal Spectrum by Interpolation Between Adjacent~Spectra}

At the previous stage, we used the optimization procedure to find
in the 4-parameter set of reddened synthetic spectra the spectrum
with the lowest deviation from that of the program star in the
LAMOST catalog. However, we can try to improve this ``optimal''
spectrum by means of interpolation between the selected spectrum
and its neighbors with the closest~parameters.

We used an iteration procedure for finding the optimal combination of
two adjacent reddened synthetic spectra by one of their four
parameters, to~obtain the minimum deviation. The~interpolation was
performed using the piecewise linear technique:
\begin{equation}
	F(\alpha,i,\lambda,\text{\textbf{P}}) 
 =
	\frac{(1-\alpha)F(P_i^{(opt)\pm1}) - \alpha F(P_i^{(opt)}) }{ P_i^{(opt)\pm1} - P_i^{(opt)} }\,,
\end{equation}
where $\alpha$~is the interpolation parameter
$(0\leq\alpha\leq1)$; $\mathbf{P}$~are the four parameters
describing reddened synthetic spectra;  $i$~is the number of the
parameter used to perform interpolation (1---$T_{\rm eff}$,
2---$\log g$, 3---$[\textrm{Fe/H}]$, 4---$A_V$), and
${(opt)\pm1}$ means that the interpolation can be performed
toward the larger optimal parameter, ``$+$'', as~well as toward
the smaller parameter, ``$-$''.

However, our modeling shows that linear interpolation of this
kind does not give any considerable reduction in the deviation.
Probably, this is the result of rather large fluctuations of real
stellar spectra compared to their computed synthetic~analogs. 

In this work, the~refinement of the optimal spectrum by interpolation between adjacent spectra was not~used.

\section{Presentation of Results of Comparison Between Synthetic and LAMOST Spectra for Single and Binary~Stars}
    \label{sec:results}

\begin{figure}[H] 
\includegraphics[width=0.65\textwidth]{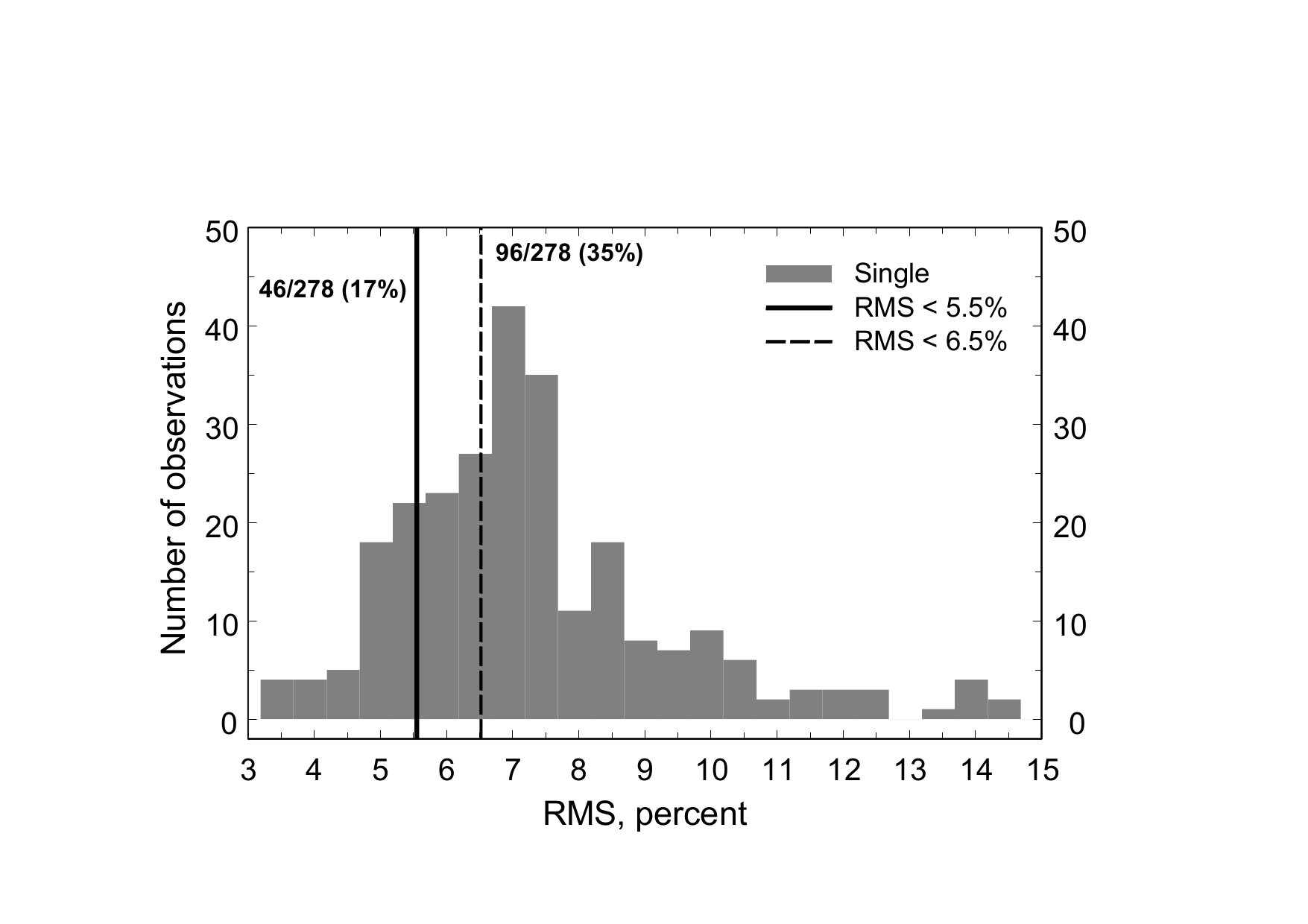}\\ 
\includegraphics[width=0.65\textwidth]{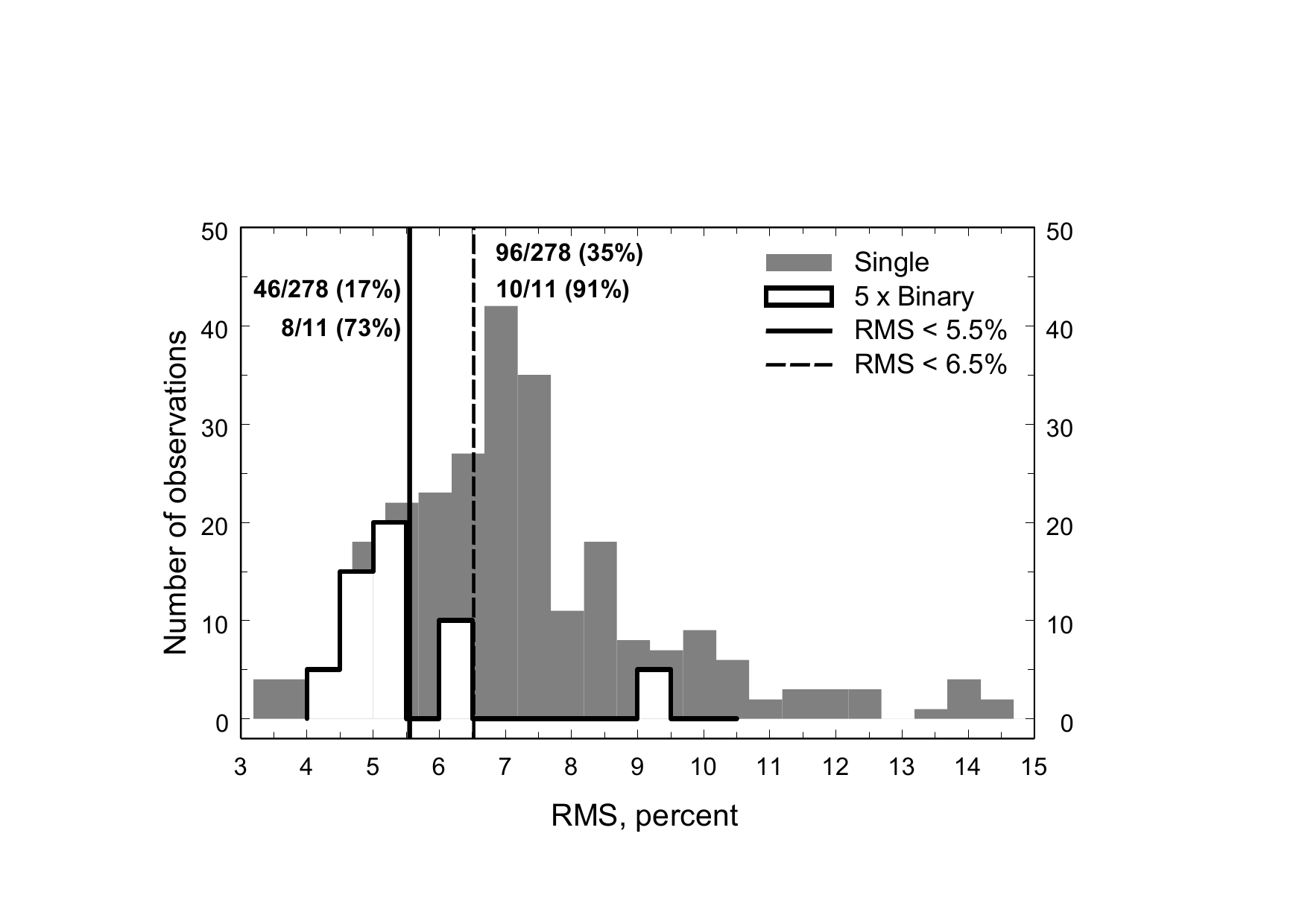} 
\caption{
Upper panel---distribution of 278 observations of 166 single stars from
the Main\_Single list over rms residual deviations. Vertical lines indicate
characteristic levels of rms = 5.5\% and 6.5\%.
Lower panel~-- same as in the upper panel, plus the histogram for
stars from SB9. The actual height of histogram columns for binaries is multiplied
by a factor of five for visibility.
\label{fig:res1}}
\end{figure}

For a quantitative evaluation of results of comparison between observed and optimized synthetic
spectra, we applied the following proximity criterion: 
\begin{equation}
	\sigma =
	\frac{ \sqrt{ \frac{1}{N}  \sum\limits_{i=1}^{N}  \left( F(\lambda_i)^{(opt)} - F(\lambda_i)^{\mathrm(LAMOST)} \right)^2 } }{ \frac{1}{N} \sum\limits_{i=1}^{N} F(\lambda_i)^{\mathrm(LAMOST)} }\,.
\end{equation}
The value of  $\sigma$~is the dimensionless ratio of  the rms deviation and its mean
value. To~make the discussion more transparent, we  express $\sigma$ in percent.
The algorithm we developed was applied to the compiled test lists
of stars. The~upper panel of Figure~\ref{fig:res1} shows the histogram of the
distribution of 278 spectra of 166 single stars from the Main\_Single
list (see Section~\ref{sec:singles}) over rms deviations between the observed LAMOST spectrum and
the corresponding optimal synthetic~spectrum.

The interval of rms $<$ 15\%, displayed in Figure~\ref{fig:res1},
contains 255 spectra of single stars. In~addition, there are 23 spectra with rms $>$ 15\% not
shown in the figure. Partially, large deviations are due to the
technical flaws of these spectra (lacking observations for a part
of the recorded LAMOST wavelength range, outliers, etc.), and 
partially, they are due to~the individual peculiarities of stellar spectra. 
The~right
edge of the diagram (rms = 15\%) was selected for clear
demonstration of the~distribution.

Since, according to the method we developed, binary stars should
have significantly larger rms deviations compared to single stars,
we marked in the histogram  two characteristic levels of rms deviations:
rms = 5.5\% and 6.5\%. In~the main list of single
stars, 46~spectra out of 278 (17\%) have rms $<$ 5.5\%;
96 spectra of 278 (35\%) have rms $<$ 6.5\%.

The lower panel of Figure~\ref{fig:res1} displays the 
histogram for the same single stars as in the upper panel and
the histogram for bona fide binary stars from the SB9 catalog. 
The~latter were processed in the same way as the LAMOST stars.
For a more clear presentation, we artificially increased
the actual height of the histogram columns for binaries by a factor of~five.

For all binaries, rms $<15$\%, so, they are visible in the
diagram. The~distribution of binary stars over the rms deviations
is more compact than for single stars. 
In the list of binary stars
from the SB9 catalog, rms $<$ 5.5\% have 8 spectra out of 11 (73\%), and
rms $<$ 6.5\% have 10 spectra out of 11 (91\%). 
Thus, the~rms deviations for binary stars from the SB9 catalog do not
stand out by their rms deviations compared to the single stars from the main list of
objects from  the LAMOST~catalog.

We additionally process the LAMOST spectra for two lists of stars identified in the Gaia~DR3 catalog. 
The first list, Gaia\_Single, contains stars that have no indications of unresolved binarity in their 
astrometric and photometric characteristics in Gaia~DR3 according to~\citet{2021A&A...649A...2L}, 
and they are most likely single stars. 
The procedure for obtaining this list of stars is presented in Section~\ref{sec:Gaia-single}. 
The Gaia\_Single list contains 6220~stars.

The second list, Gaia\_Binary, contains stars that do not have significant indications of 
unresolved binarity according to the Gaia~DR3 catalog. The~procedure for obtaining this 
list of stars is given in Section~\ref{sec:Gaia-binary} above. The~Gaia\_Binary list 
contains 2278 stars. The~stars in this list are most likely unresolved~binaries.

\begin{figure}[H] 
\includegraphics[width=0.65\textwidth]{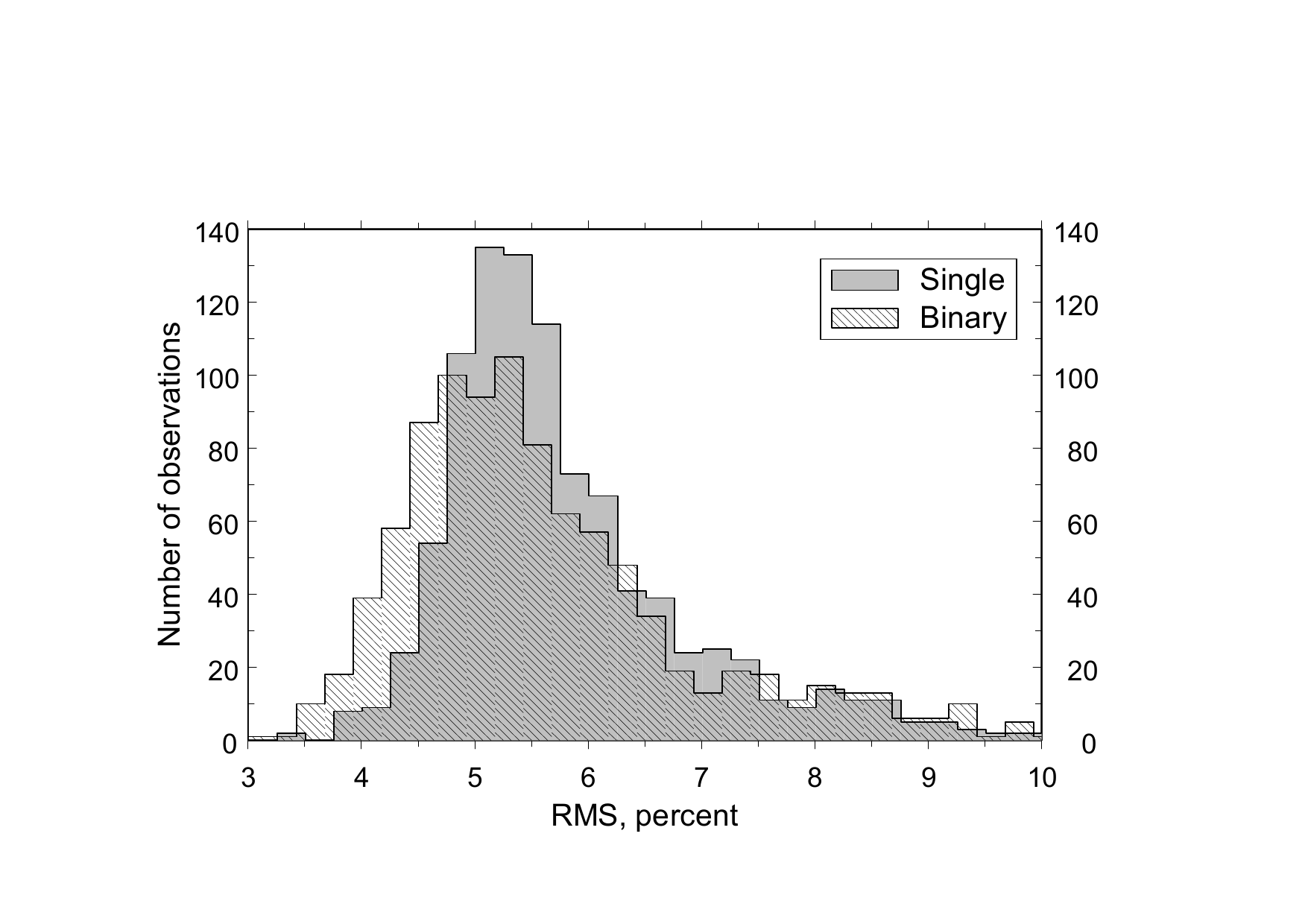} 
\caption{
Distribution of 1000 putative single stars (gray histogram) 
and 1000 putative unresolved binaries (shaded histogram) 
from the Gaia DR3 catalog. Bin width 0.25.}
\label{fig:Gaia-res}
\end{figure}

Figure~\ref{fig:Gaia-res} shows histograms of rms distributions of stars from the 
Gaia\_Single and Gaia\_Binary lists. 
For~each histogram, 1000 stars from each list were randomly selected. 
The~histogram bin width is 0.25. 
The~distributions 
are almost identical: the average values are $\langle rms_{Singl}\rangle = 6.3$ and 
$\langle rms_{Bin}\rangle = 6.2$. 
The~distribution of unresolved binary stars 
is slightly wider: standard deviations are $\sigma_{Singl} = 3.4$ and $\sigma_{Bin} = 4.2$. 
Beyond the right edge of the figure ($rms>10.0$) there are 60 single and 63 
unresolved binary stars out of~1000.

Thus, the~binary and single stars from the Gaia~DR3 catalog 
differ only slightly and do not stand out in their rms against the background 
of the Main\_Single stars considered~above.

\unskip

\section{Conclusions}\label{sec:6}
\label{sec:conclusion}

We have suggested a method to find binary stars among LAMOST sources, based on the comparison between
the continuum of the stellar spectrum of the  LAMOST catalog and a
synthetic spectrum, with~interstellar reddening
taken into account. It has been demonstrated that this method is not effective enough in its version
described above and is apparently not able to distinguish between
single and binary stars in the LAMOST~catalog.

The possible directions of further studies are as follows:

\begin{enumerate}
\item Compilation of a special version of the CKL synthetic spectra
library, with~wavelengths exactly coinciding with those in the LAMOST
spectra. 
Using such spectra would eliminate procedures of
smoothing and interpolation, leading to lower rms deviations
between the observed and synthetic spectra and to better
sensitivity of the method. However, computation of stellar spectra 
library is a stand-alone~problem.

\item Switching to a method based on a comparison of intensities for
pairs of spectral lines or intensities for individual lines. The~first of these methods is currently the most widely used approach
for high-accuracy spectral classification of stars. The~second
approach is used for spectral classification of stars in the
LAMOST project proper. However, the~LAMOST survey provides
low-resolution stellar spectra, and~such methods can also turn out
to be not sensitive to stellar~binarity.
\end{enumerate}

{\bf Author contributions:} Conceptualization, S.N., T.K., M.O. and Z.G.;
methodology, T.K., S.S., L.A., K.D. and Z.G.; software, K.P. and S.S.; validation, P.M.;
investigation, Y.L.; resources, L.A., Z.J., L.Y., S.Y. and Z.G.;
data curation, T.K., L.A., Z.J., L.Y., S.Y. and Z.G.; writing-original draft preparation, S.N.;
writing-review and editing, P.M., T.K., S.N., L.A., K.D., Z.J., L.Y., K.P., M.O., S.Y., S.S., Y.L. and Z.G.;
visualization, S.S., K.P. and T.K.; supervision, P.M. and T.K.; project administration, P.M.
All authors have read and agreed to the published version of the manuscript.

{\bf Funding:} 
G.Z. is supported by the National Natural Science Foundation of China (NSFC) under
grant No. 11988101 and by the National Key R\&D Program of China No. 2024YFA1611900.
K.T. is supported by the NSFC under grant Nos. 12261141689, 12090044, and~12090040.
Y.L. is supported by the NSFC under grant No. 12073044.
O.M. thanks the CAS President's International Fellowship Initiative (PIFI).
Guoshoujing Telescope (the Large Sky Area Multi-Object Fiber Spectroscopic Telescope LAMOST) is a National Major Scientific Project
built by the Chinese Academy of Sciences. Funding for the project has been provided by the National Development and Reform Commission. LAMOST
is operated and managed by the National Astronomical Observatories, Chinese Academy of Sciences.
The study was conducted under the state assignment of Lomonosov Moscow State University.

{\bf Acknowledgments:}  This research has made use of the VizieR catalog access tool, CDS, Strasbourg, France. This work has made use of data from the European Space Agency (ESA) mission Gaia 
(\url{https://www.cosmos.esa.int/gaia} (accessed on 28.07.2025), 
processed by the Gaia Data Processing and Analysis Consortium 
(DPAC, \url{https://www.cosmos.esa.int/web/gaia/dpac/consortium}, accessed on 28.07.2025). 
Funding for the DPAC has been provided by national institutions, in~particular the institutions participating in the Gaia Multilateral Agreement.
The use of TOPCAT, an~interactive graphical viewer and editor for tabular data \cite{2005ASPC..347...29T}, is~acknowledged.
We are grateful to our anonymous reviewers whose constructive comments greatly helped us to improve the paper.

{\bf Conflicts of interests:} The authors declare no conflicts of~interest.

\begin{adjustwidth}{-0cm}{0cm}
\bibliographystyle{apacite}


\end{adjustwidth}

\end{document}